\documentclass[aps,prx,amsmath,amssymb,
showpacs,
reprint,%
longbibliography
]{revtex4-1}

\usepackage[T1]{fontenc}
\usepackage{graphicx}

\usepackage{amsmath}
\usepackage{times}

\newcommand{\Rb}{\mathbf{R}}
\newcommand{\qb}{\mathbf{q}}

\newcommand{\vb}{\mathbf{v}}
\newcommand{\eb}{\mathbf{e}}

\newcommand{\inner}[2]{\langle#1|#2\rangle}
\newcommand{\icm}{cm$^{-1}$}
\newcommand{\gdos}{GDOS}

\begin{document}

\title{Efficient method for calculating Raman spectra of solids with impurities and alloys and its application to two-dimensional transition metal dichalcogenides}

\author{Arsalan Hashemi$^1$}
\author{Arkady V. Krasheninnikov$^{1,2}$}
\author{Martti Puska$^1$}
\author{Hannu-Pekka Komsa$^1$}

\affiliation{
$^1$Department of Applied Physics, Aalto University,
P.O. Box 11100, 00076 Aalto, Finland
}
\affiliation{
$^2$Helmholtz-Zentrum Dresden-Rossendorf,
Institute of Ion Beam Physics and Materials Research, 01328 Dresden, Germany
}

\date{\today}

\begin{abstract}
Raman spectroscopy is a widely used, powerful, and nondestructive tool
for studying the vibrational properties of bulk and low-dimensional materials.
Raman spectra can be simulated using first-principles methods,
but due to the high computational cost calculations are
usually limited only to fairly small unit cells,
which makes it difficult to carry out simulations for alloys and defects.
Here, we develop an efficient method for simulating Raman spectra of alloys,
benchmark it against full density-functional theory calculations,
and apply it to several alloys of two-dimensional transition metal dichalcogenides.
In this method, the Raman tensor for the supercell mode is constructed 
by summing up the 
Raman tensors of the pristine system 
weighted by the projections of the supercell vibrational modes
to those of the pristine system.
This approach is not limited to 2D materials and should be applicable to 
any crystalline solids with defects and impurities.
To efficiently evaluate vibrational modes of very large supercells,
we adopt mass approximation, although it is limited to chemically
and structurally similar atomic substitutions.
To benchmark our method, we first apply it to Mo$_x$W$_{(1-x)}$S$_2$ monolayer in the H-phase, where several experimental reports are available for comparison.
Second, we consider Mo$_x$W$_{(1-x)}$Te$_2$ in the T'-phase, which has been
proposed to be 2D topological insulator, but where experimental results
for the monolayer alloy are still missing. 
We show that the projection scheme also provides a powerful tool 
for analyzing the origin of the alloy Raman-active modes 
in terms of the parent system eigenmodes.
Finally, we examine the trends in characteristic Raman signatures for 
dilute concentrations of impurities in MoS$_2$.
\end{abstract}
\maketitle
\section{Introduction}

Two-dimensional (2D) materials have been extensively studied for applications
in optoelectronics, thermoelectrics, sensing, catalysis, etc.
While the catalogue of available 2D materials is vast \cite{Nicolosi13_Sci,Lebegue13_PRX,Mounet18_NNano},
it may be difficult to find a material that perfectly suits the desired
specifications. In such cases, alloying can be used to further tune
the material properties.
Taking the transition metal dichalcogenide (TMD) family of 2D materials as an example, alloying the prototypical member MoS$_2$ with WS$_2$ or MoSe$_2$
leads to straightforward modification of
electrical conductivity \cite{Revolinsky64_JAP,Srivastava97_SM},
band gap and band edges
\cite{Chen13_ACSNano,Komsa12_JPCL,Kang13_JAP,Li14_JACS,Mann14_AM,Rigosi16_PRB},
and spin-orbit splitting \cite{Wang15_NComm}.
More interestingly, alloying can even provide properties that were not present 
in the constituent phases.
For instance, alloying can lead to dramatic reduction of the thermal conductivity
\cite{Gu16_PRB,Qian18_APL} or passivation of defect levels \cite{Huang15_PRL,Yao16_ACSAMI}.
The beneficial role of alloying has already been demonstrated in 
few applications: the response characteristics of
(Mo,W)S$_2$-based photodetector \cite{Yao16_ACSAMI}
and the catalytic activities of Mo(S,Se)$_2$ 
alloys \cite{Kiran14_Nanos,Wang15_AM} were found to be 
better than in their parents.

Among TMD alloys,
a particularly curious case is (Mo,W)Te$_2$ alloy, since MoTe$_2$
is more stable in the H-phase and WTe$_2$ is more stable in the T'-phase, although the energy differences between the two phases are small for
both parent materials and, in fact, MoTe$_2$ can also be grown in the T'-phase.
The phase tunability is particularly interesting for these materials, as they have drastically different electronic properties in different phases.
In the H-phase, these materials are semiconductors, while in the T'-phase 
they are semimetals or topological insulators depending on the number of layers
\cite{Cazalilla14_PRL,Qian14_Sci,Sun15_PRB,Huang16_NMat}.
Due to similar energies, coexistence of H/T phase regions has
been predicted in Ref.\ \onlinecite{Duerloo16_ACSNano}, 
and it was also proposed that the H/T'-transition in (Mo,W)Te$_2$
could be promoted by gating \cite{Zhang16_ACSNano}.
Moreover, 2D ferroelectricity was recently demonstrated in T'-WTe$_2$ 
even in the monolayer limit \cite{Fei18_Nat}.

Raman spectroscopy is an important and versatile tool for
characterizing the composition of 2D alloys and assessing
their overall crystal quality but it is not always straightforward
to assign new peaks (as compared to the spectrum of the parent systems)
to the structural features from which they originate from.
Several TMD alloys have already been extensively studied in the literature
by Raman spectroscopy 
providing datasets covering a full composition range in many alloy systems
such as
(Mo,W)S$_2$ \cite{Chen14_Nanos,Liu14_Nanos,Park18_ACSNano}
(Mo,W)Se$_2$ \cite{Tongay14_APL,Zhang14_ACSNano}
Mo(S,Se)$_2$ \cite{Mann14_AM,Feng15_ACSNano,Su14_Small,Li14_JACS}
Re(S,Se)$_2$ \cite{Wen17_Small}.
For bulk alloys, similar studies are also done \cite{Dumcenco10_JAC}
and for T'-(Mo,W)Te$_2$ we are only aware of bulk alloy 
studies \cite{Oliver17_2DM,Lv17_SRep}, but not of monolayer alloys.

There also exists a lot of computational studies for the Raman spectra for pristine, constituent phases \cite{MolinaSanchez11_PRB,Zhang15_CSR,Saito16_JPCM},
and even few reports for defective MoS$_2$ \cite{Parkin16_ACSNano,Bae17_PRA}.
Despite the importance of Raman spectroscopy in understanding the alloy composition
and the structural order, computational studies for alloys are missing.
The reason is that, within the conventional computational approach,
these calculations are computationally
significantly more challenging due to the larger supercells involved
and the dramatic scaling of the computational cost with the supercell size.
When the maximum computationally feasible supercells sizes are often 3$\times$3
or at maximum 6$\times$6 primitive cells it is clear that
(i) the impurity/defect concentration is necessarily high, and
(ii) the defects are ordered and thus the simulated spectra
for a given alloy are unlikely to 
correctly mimic that of the randomly distributed system.
These issues need to be tackled before computational Raman spectra for alloys can be calculated in a way that can be reliably compared to experiments and even holds predictive power.

In this paper, we propose a computational method to simulate
Raman spectra of alloys using large supercells.
The method relies on the projection of the vibrational eigenvectors of the supercell 
to those of the primitive cell, which are then used to weight the 
Raman tensors of the pristine system.
When the lattice constants and the bonding chemistry in the two components
are similar, as is the case in the systems considered here, the supercell eigenvectors
can be efficiently solved using the mass approximation.
We benchmark our method both towards the full DFT approach in small supercells
as well as experimental results.
We first apply our method with the (Mo,W)S$_2$ alloy, for which extensive experimental results are available.
We analyze the modes and, in particular, try to distinguish between the
one-mode and two-mode behavior, and visualize the eigenmodes that contribute to the most prominent Raman peaks.
Next, we consider T'-phase MoWTe$_2$, which is much more involved due to the lower symmetry, larger supercell, and (semi-)metallic electronic structure,
while a mass approximation is expected to hold equally well. 
Finally, we consider dilute concentrations of impurities in MoS$_2$, both in the
Mo site and in the chalcogen site, and look for characteristic Raman signatures.

\section{Methods}

\subsection{Theoretical framework}

As mentioned in the introduction, first-principles Raman calculations
for large unit cells are computationally challenging.
They involve two steps: (i) determination of the vibrational modes
of the system and (ii) calculation of the Raman activity for each mode.

In step (i), the vibrational modes (eigenmodes) are solutions to
\begin{align} \label{eq:motion}
    M_k \omega^2 \vb(k0) &= \sum_{k',l} \Phi(k'l,k0) \vb(k'l) \\
            &= \sum_{k',l} \Phi(k'l,k0) \exp(-i\qb\cdot \Rb_l) \vb(k'0).
\end{align}
where $\vb(kl)$ are the eigenvectors for the displacement of atom $k$ with mass $M_k$
located in cell $l$ specified by the lattice vector $\Rb_l$.
The elements of 
force constant (FC) matrix $\Phi$ are defined by
the change of potential energy, $U$, with respect to the atomic displacements
$$
\Phi_{\alpha\beta}(k'l,k0) = \frac{\partial^2 U}{\partial u_\alpha(k'l)\partial u_\beta(k0)} .
$$
Above, $u_{\alpha}(kl)$ denotes the displacement of the $k$th atom in
the $l$th unit cell in the cartesian direction $\alpha$.
Constructing the force constant matrix in the case of alloys, without any symmetry,
essentially requires performing $3N$ DFT total energy calculations
in which each of the $N$ atoms is displaced in each of the three cartesian directions.

In step (ii), the Raman intensity can be written as
\begin{equation}\label{eq:Rint}
I \sim \rvert \eb_s \cdot R \cdot \eb_i \rvert^2 
\end{equation}
where $\eb_i$ and $\eb_s$ denote the polarization vectors of the incident and scattered light
and $R$ is the Raman tensor.
In the case of nonresonant first-order Raman scattering,
it is obtained from the change of polarizability $\chi$ 
with respect to the phonon eigenvectors $\vb$,
and in first-principles calculations
it can be evaluated by using the macroscopic dielectric constant
$\varepsilon_{\rm mac}$ as
\begin{equation}\label{eq:Rpol}
R \sim \frac{\partial \chi}{\partial \vb} 
  = \frac{\partial \varepsilon_{\rm mac}}{\partial \vb}.
\end{equation}
This derivative needs to be evaluated at both the positive and negative displacements
for each of the $3N$ eigenvectors $\vb$, yielding a total of $6N$ calculations. 
Moreover, in spite of different approaches, evaluating $\varepsilon_{\rm mac}$ is generally 
significantly more time-consuming than DFT total energy calculations.

While step (ii) takes more time, already step (i) becomes challenging in large low-symmetry systems.
In case of MoS$_2$, the limitations are currently at around 10$\times$10
supercell for step (i) and 6$\times$6 supercell for step (ii).
In order to properly account for the random distribution of atoms and the
resulting broadening of the spectra, large supercells or averaging over several
configurations is required.
Herein, we adopt two approximations to tackle each of these issues:
a mass-approximation for step (i) and 
projection to the primitive cell Raman-active eigenmodes for step (ii).

In the mass-approximation (MA), only masses are changed in Eq.\ \ref{eq:motion},
whereas the force-constant matrix remains untouched \cite{Baroni90_PRL,Menendez}.
Naturally, this can only be applied in cases where the nature of the bonding and
the atomic structure remain very similar,
such as for instance Al$_x$Ga$_{1-x}$As \cite{Baroni90_PRL}.

Due to the small momentum of photons commonly used in Raman spectroscopy, 
and especially in non-resonant Raman where the photon energy needs to be less than the band gap,
first-order Raman scattering can only involve single phonon near $q=0$.
For pristine materials the $q=0$ phonons are trivially obtained
as the $\Gamma$-point solutions of Eq.\ \ref{eq:motion}
in the primitive cell (PC).
If we consider a supercell (SC) of pristine material, 
the $\Gamma$-point contains several modes from the folding of the phonon bands.
In an explicit calculation of Raman intensities using Eq. \ref{eq:Rpol}
the intensities of the folded modes will be zero and thus the Raman spectra
remains the same.
Alternatively, the folded modes in the supercell $\Gamma$-point
could be unfolded back to the primitive cell Brillouin zone (BZ) through projection to plane waves $g(\qb) = \exp(i\qb \cdot \Rb)$, where
$\qb$ corresponds to one of the PC q-points that fold into the $\Gamma$-point of the SC.
Adopting the notation where $\vb^{\rm SC}(kl)$ refers to the $l$th
primitive cell within the supercell and $k$ indexes the atoms in the unit cell,
the projection is written out as
\begin{equation}\label{eq:proj}
 \inner{g(\qb)}{\vb^{SC}}_{\alpha,k} = 
   \sum_l \exp(-i\qb \cdot \Rb_{l}) v_{\alpha}^{SC}(kl).
\end{equation}
While we could use this equation to unfold to any $\qb$, we are here primarily
interested in the $\Gamma$-point, which fortuitously also yields a particularly simple expression since the exponent
in Eq. \ref{eq:proj} is always unity and thus one ends up with a straightforward sum
over the eigenvectors.
The total $\Gamma$-point weight can be obtained by taking the square of the projections 
and summing up over $k$ and $\alpha$. Finally, we sum up over all the SC states $i$
with frequency $\omega_i$ to obtain $\Gamma$-point weighted density-of-states
\begin{equation}\label{eq:proj2}
   n(\omega) = \sum_i \sum_{\alpha,k} \rvert \inner{g(\qb)}{\vb^{SC,i}}_{\alpha,k} \rvert^2
   \delta(\omega-\omega_i)
\end{equation}
which we here denote as \gdos{}.
Since each mode in pristine supercell has non-zero weight in only a single q-point 
in the PC BZ, the true $q=0$ modes can easily be found.
In alloys or defective systems, where the translational symmetry is broken,
the unfolding/projection procedure still works, but leads to each SC mode having 
contributions from q-points throughout the PC BZ with different weights.
This type of unfolding procedures have already been used in the past
to analyze both the electronic and phonon band structures of alloys
\cite{Allen2013,Zheng2016, Huang2014, Gordienko17_PSSB}.

Baroni et al. found that the GDOS of the primitive cell can be 
used to closely approximate the Raman spectra \cite{Baroni90_PRL} in alloys.
The modes which were inactive due to momentum-conservation law can gain weight at $q=0$ 
and start to show up in the Raman spectra and, vice versa, the modes that were originally purely
$q=0$ modes can leak weight to other q-points and thereby lose Raman intensity.
Such analysis is straightforward when the frequencies of Raman-active and
-inactive modes are clearly separated.
If they are close, it is no longer clear which part of the GDOS would be Raman-active.
To solve this issue, we here propose to project the SC modes not to plane waves
but to PC eigenmodes at the $\Gamma$-point.
That is,
adopting the same notation for $\vb^{SC}(kl)$ as above,
\begin{equation}\label{eq:projeig}
    w_{ij} = \inner{\vb^{PC,i}}{\vb^{SC,j}}  =
    \sum_{\alpha,k,l} v_{\alpha}^{PC,i}(k0) v_{\alpha}^{SC,j}(kl).
\end{equation}
Here, due to the mass-approximation, the atoms are in the same positions both in the alloy
and in the pristine cells.
However, it appears to work well also with the DFT relaxed structures.
Since the projection is to PC modes at the $\Gamma$-point, we simultaneously obtain the $\Gamma$-point projection (or unfolding).
We note, that the summation of projections $w_{ij}^2$ over 
all $k$ and $\alpha$ yields the same
GDOS as via the plane wave projections (Eq.\ \ref{eq:proj2}),
since both constitute a complete basis set.
The Raman tensor of the SC mode is obtained by multiplying the PC mode projection
by the respective Raman tensors from the pristine system, i.e.,
\begin{equation}\label{eq:Rsum}
R^{\rm SC,j} = \sum_{i} w_{ij} R^{\rm PC,i}
\end{equation}
where the sum goes over PC modes $i$ and clearly only Raman-active modes contribute.
Finally,
the Raman intensity of the SC mode $j$
is obtained using Eq.\ \ref{eq:Rint} which yields
\begin{align}
I^{\rm SC,j} &\sim \rvert \eb_s \cdot R^{\rm SC,j} \cdot \eb_i \rvert^2  \label{eq:RGDOS} \\
  &= \sum_i w_{ij}^2 \rvert \eb_s \cdot R^{\rm PC,i} \cdot \eb_i \rvert^2 \nonumber \\
  &+ \sum_{i\neq k} ( \eb_s \cdot w_{ij} R^{\rm PC,i} \cdot \eb_i )^* ( \eb_s \cdot w_{kj} R^{\rm PC,k} \cdot \eb_i ) \label{eq:RGDOS2} \\
  &\approx \sum_i w_{ij}^2 I^{PC,i}. \label{eq:RGDOS3}
\end{align}
Squaring the sum over PC modes leads to $i=k$ and $i\neq k$ terms,
which have been separated in the second step.
These cross terms can be important if the PC mode has appreciable weight arising from several PC modes. 
In the last step, we have assumed that they are negligible. 
While indeed not always a good assumption, the advantage is that 
we can now sum over intensities rather than Raman tensors.
This is useful because we could then, e.g.,
use experimentally determined intensities instead of the calculated ones.
We denote the total Raman intensity weighted GDOS as RGDOS.
When the contributions from each mode to the total
Raman spectra are shown in the Results section, these correspond only to
the first term in Eq.\ \ref{eq:RGDOS2}.
We note, that in some previous works the Raman tensor in alloy/defective supercells
has been decomposed using the Raman tensors of different symmetries of the pristine host
for the analysis purposes \cite{Ikeda17_PRB,Qian18_Langm}.
Here, we essentially proceed in the opposite direction in order
to construct the final Raman tensor.
Moreover, our approach is in principle more general as it can distinguish 
between different modes of the same symmetry.

To sum up, the main ingredients of the method lie in the projection of supercell
vibrational eigenmodes to the pristine system eigenmodes (Eq.\ \ref{eq:projeig}),
and using those projections as weights when summing up over the primitive cell
Raman tensors (Eq.\ \ref{eq:Rsum}).
The general applicability of our method is mostly limited by the eigenmode projection,
which essentially requires that there needs to be a reasonable mapping between the
atomic structures of the non-pristine and pristine systems.
Extension of the method to simulate second-order non-resonant scattering 
should be fairly straightforward.
To simulate resonant Raman scattering, in principle one can just plug the resonant 
Raman tensors to Eq.\ \ref{eq:Rsum}. In practice, the modifications of the electronic
structure need to be also carefully considered, the details of which strongly depend
on the system.

\subsection{Computational details and benchmarking}

All first-principles calculations are carried out with VASP \cite{VASP}.
Exchange-correlation contributions are treated with 
the PBEsol functional \cite{PBEsol}.
A plane wave basis with a cutoff energy of 550 eV is employed to
represent the electronic wave functions.
The geometry optimization continues until the energy differences and ionic forces are converged to less than $10^{-6}$ eV and 1 meV/\AA, respectively.
The first Brillouin zone of primitive cell is sampled by a 12$\times$12 mesh 
for H-MoS$_2$/WS$_2$ and by a 12$\times$24 mesh for T'-MoTe$_2$/WTe$_2$,
and, changing in proportion to the supercell size N.
The polarizability tensors for Raman calculations
are determined within the framework of the finite
displacement method \cite{Raman_unpolarized}.
The phonon spectra are assessed using the PHONOPY code \cite{PHONOPY}
using 6$\times$6 supercell for MoS$_2$/WS$_2$ and
4$\times$4 supercell for MoTe$_2$/WTe$_2$.
The Raman intensity is calculated as an average over
the XX and XY configurations for the light polarization ($\eb_i$$\eb_s$).

\begin{figure}[!ht]
\begin{center}
  \includegraphics[width=8cm]{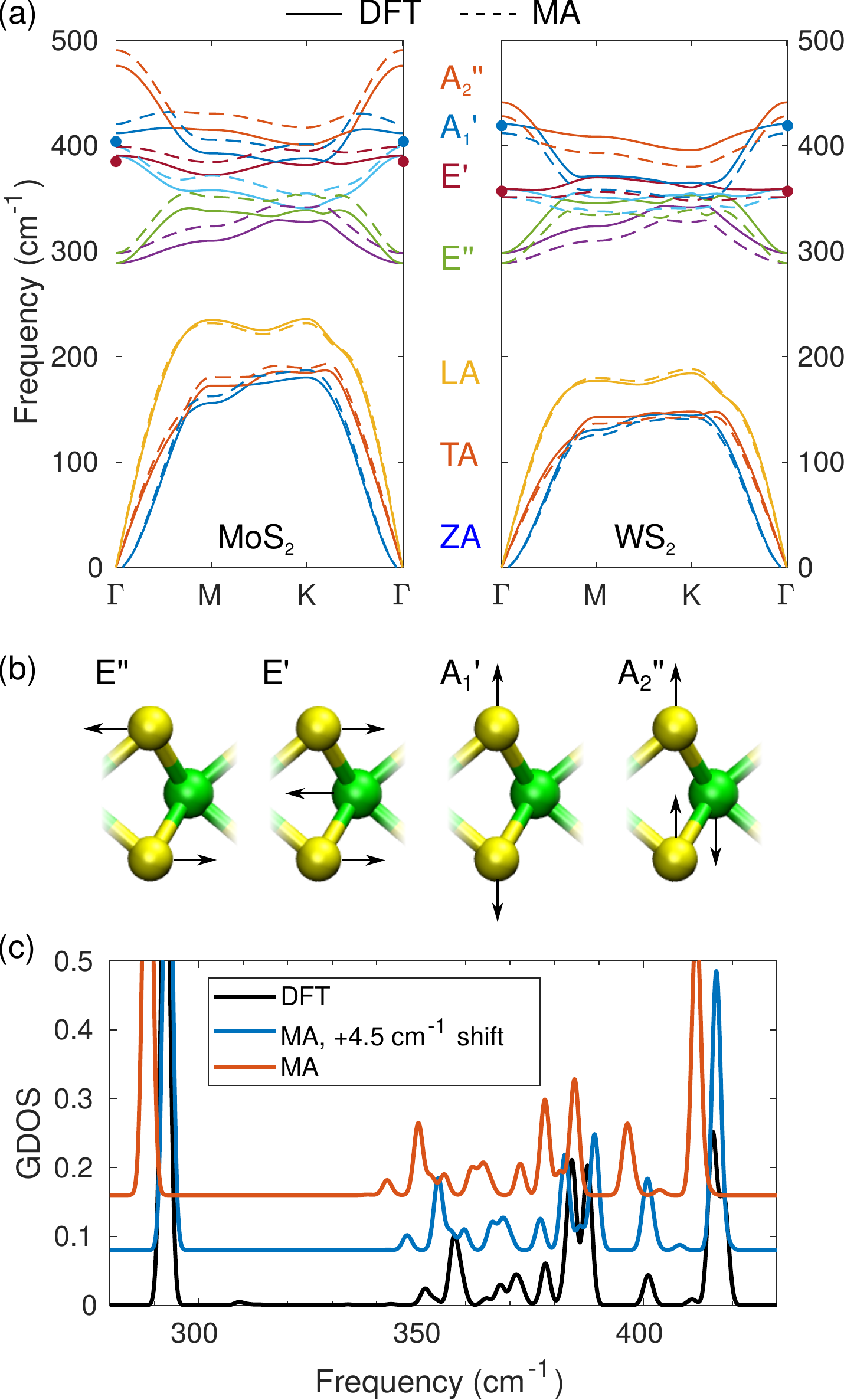}
\end{center}
\caption{\label{fig:valid}
(a) Phonon dispersion curves of pristine MoS$_2$ (left) and WS$_2$ (right)
calculated either self-consistently using DFT (solid lines) or
using the mass-approximation (dashed lines).
Dots denote experimental values obtained
from Raman spectroscopy \cite{Zhang15_CSR,Livneh15_2DM}.
(b) Schematic representation of Mo (green) and S (yellow) atoms vibrations
in different optical phonon modes.
(c) GDOS from 3$\times$3 SQS of Mo$_{0.56}$W$_{0.44}$S$_2$, either calculated
fully with DFT or using the mass approximation.
}
\end{figure}

We start by benchmarking our computational scheme
with respect to the mass-approximation.
We show in Fig.\ \ref{fig:valid}(a)
the phonon dispersion curves of MoS$_2$ and WS$_2$ calculated with DFT and 
the mass-approximated versions (i.e., using the MoS$_2$ FC matrix
but substituting the mass of Mo by that of W and vice versa).
The dispersions of the bands are captured very well with MA as are the acoustic mode frequencies.
There is a nearly constant downshift of the optical mode frequencies 
of WS$_2$ by about 10 cm$^{-1}$ with respect to self-consistent WS$_2$ calculation,
and vice versa an upshift in MoS$_2$ frequencies if using WS$_2$ FC with Mo mass,
suggesting that W-S bonds are slightly stronger than Mo-S bonds.
In the following of this work, we have chosen to use the MoS$_2$ force constants.
With this choice, when comparing to the experimental values for the 
two Raman-active modes, E$'$ and A$_1'$,
our calculated frequencies are slightly overestimated for MoS$_2$
and slightly underestimated for WS$_2$ when compared to full DFT calculation.

The effect of MA is further illustrated in Fig.\ \ref{fig:valid}(c) in the case of the (Mo,W)S$_2$ alloy supercell.
The structural models used in the alloy calculations are 
constructed using the special quasirandom structures (SQS) method \cite{Zunger1990}.
As seen in Fig.\ \ref{fig:valid}(c)
for 3$\times$3 Mo$_{0.56}$W$_{0.44}$S$_2$ SQS,
the MA frequencies are downshifted throughout the spectrum,
similar to the pristine systems.
To allow for a better comparison with the DFT results, we also show a spectrum shifted up by 4.5 cm$^{-1}$ (from the alloy composition times 10 cm$^{-1}$), 
after which the main peaks (E$'$, A$_1'$) and the high-frequency
part of the E$'$ feature (from 350 to 400 cm$^{-1}$) agree very well. 
The low-frequency part of E$''$ features has still a too low frequency, which is due
to the fact that these modes are localized to W atoms, as will be seen later.
\begin{figure}[!ht]
\begin{center}
  \includegraphics[width=8cm]{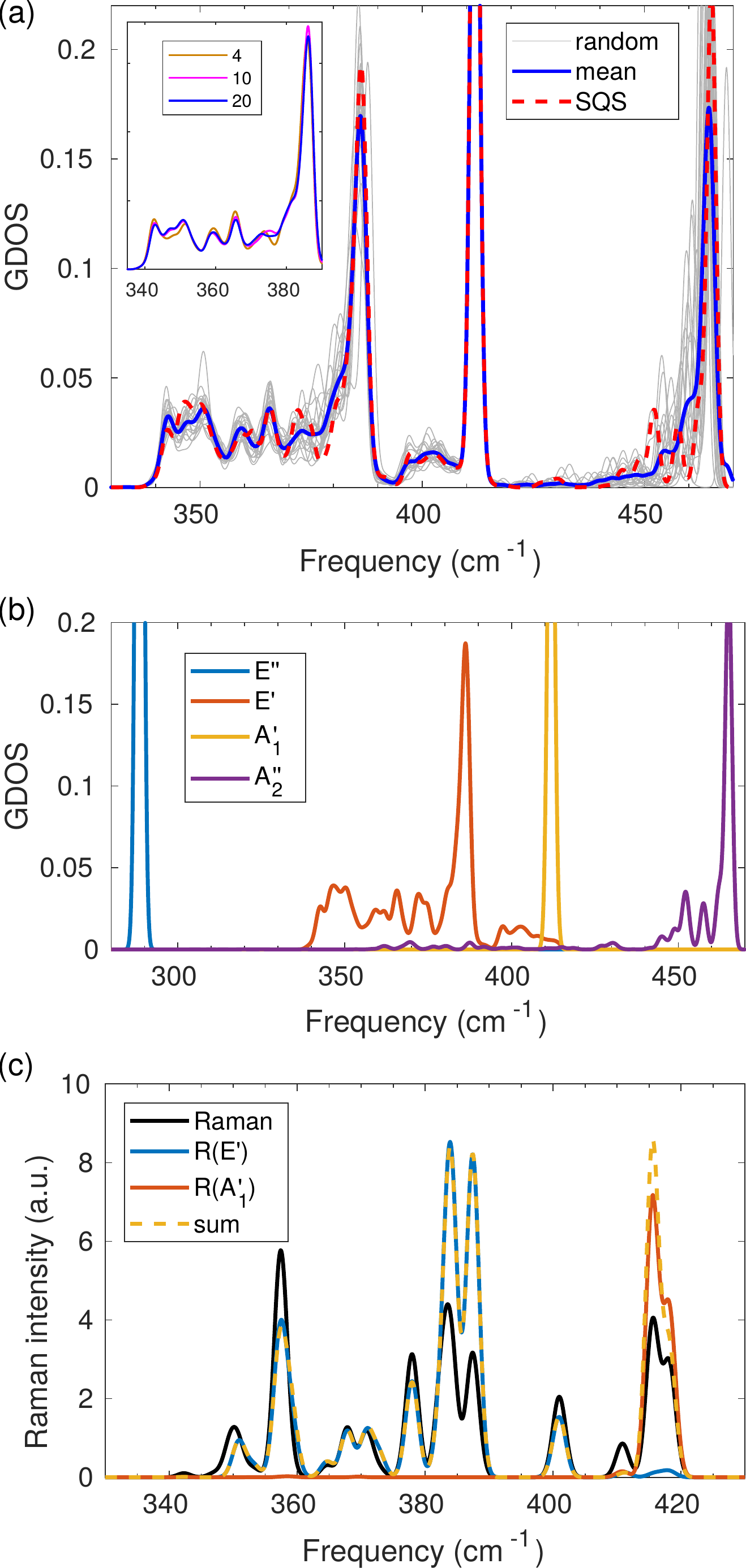}
\end{center}
\caption{\label{fig:valid2}
(a) GDOS from 20 random atomic configurations for Mo$_{0.56}$W$_{0.44}$S$_2$
(gray lines) together with their average (blue, solid line).
GDOS for the 12$\times$12 SQS supercell is also shown for comparison.
Inset: Comparison of results when the averaging is done over 4, 10, or 20 configurations.
(b) Primitive cell eigenmode projected GDOS for the SQS cell [same as in panel (a)].
(c) Raman spectra from full DFT calculation for the 3$\times$3 SQS supercell
of Mo$_{0.56}$W$_{0.44}$S$_2$ compared to its RGDOS,
and the contributions to it from the two primitive cell eigenmodes.
}
\end{figure}

Next, we inspect the importance of statistical sampling. 
We use the supercell comprising 12$\times$12 primitive cells
and 20 different random configurations (not SQS) for each composition.
Fig. \ref{fig:valid2}(a) shows the spectra from all the
20 configurations and the averaged spectra.
The large variation in the single spectra
indicates that 12$\times$12 supercell is still not quite large enough
to correctly describe the alloy with a single supercell. 
As shown in the inset, averaging over just 4 configurations yields a spectrum
that is already quite similar to that from 20 configurations.
In addition, we compare the averaged spectrum to that of
a SQS model created within the 12$\times$12 supercell.
We consider pairs up to 8 {\AA} (three effective cluster interaction (ECI) parameters) and three-body clusters up to 4 {\AA} (2 ECI).
The SQS performs better than the different random configurations, but
fails to correctly capture the smooth broadening of the main peaks,
instead yielding more spiked features.
This originates from the coarseness of the mesh of k-points that 
folds into the $\Gamma$-point in small supercell calculations.
Note, that the A$_1'$ mode is in practice
completely unaffected by the mixing, as it only
involves movement of the chalcogen atoms and the metal atoms are fixed
(see Fig. \ref{fig:valid}(b)).

Finally, we benchmark the eigenmode-projection scheme.
First, we illustrate in Fig.\ \ref{fig:valid2}(b) the eigenmode contributions 
in the case of 12$\times$12 SQS. 
In H-MoWS$_2$ alloy, the modes remain fairly separated in frequency and thus the resulting Raman spectra could be fairly safely evaluated from just the GDOS.
On the other hand, the projection scheme provides further insight in to the origin of the spectral features. 
For instance, the bump at around 400 cm$^{-1}$ originates from the E$'$ mode and not from the A$_1'$ mode.
Also, at large W concentration, the A$_2''$ features start to overlap with
the E$'$/A$_1'$ features, as will be seen in the Results section.
Moreover, we need to compare how well the approximated Raman spectra
match to explicit Raman calculations. 
For this we need to adopt a smaller system, and since this is only
for benchmarking purposes we can take a 3$\times$3 supercell,
again created using the SQS scheme.
The RGDOS captures surprisingly well all the features of the full Raman
calculation, as shown in Fig.\ \ref{fig:valid2}(c).
Especially, the peak shapes/structures are correctly reproduced, even if
some intensities differ
with the most significant discrepancy occuring near 385 cm$^{-1}$.
From the comparison of the spectra in Figs.\ \ref{fig:valid2}(b) and (c)
it is again obvious that 3$\times$3 SQS cannot describe properly
the Raman spectrum of the random alloy.

We have demonstrated that large supercells are needed to properly describe the
phonon spectra of random alloys and that RGDOS can be used to give a
good estimate of the Raman spectrum. 
While the mass approximation may produce some inaccuracies with
the peak positions, we feel that this is acceptable tradeoff for
the ability to correctly describe the random alloy.
In the following, the results for the alloys are obtained by averaging
over 20 configurations of the 12$\times$12 supercell and
using the eigenmode-projection.
In few cases, the analysis of the results is done using the SQS structure,
which results in great simplification.

\section{Results}

\subsection{H-(Mo,W)S$_2$}

\begin{figure*}[!ht]
\begin{center}
  \includegraphics[width=16cm]{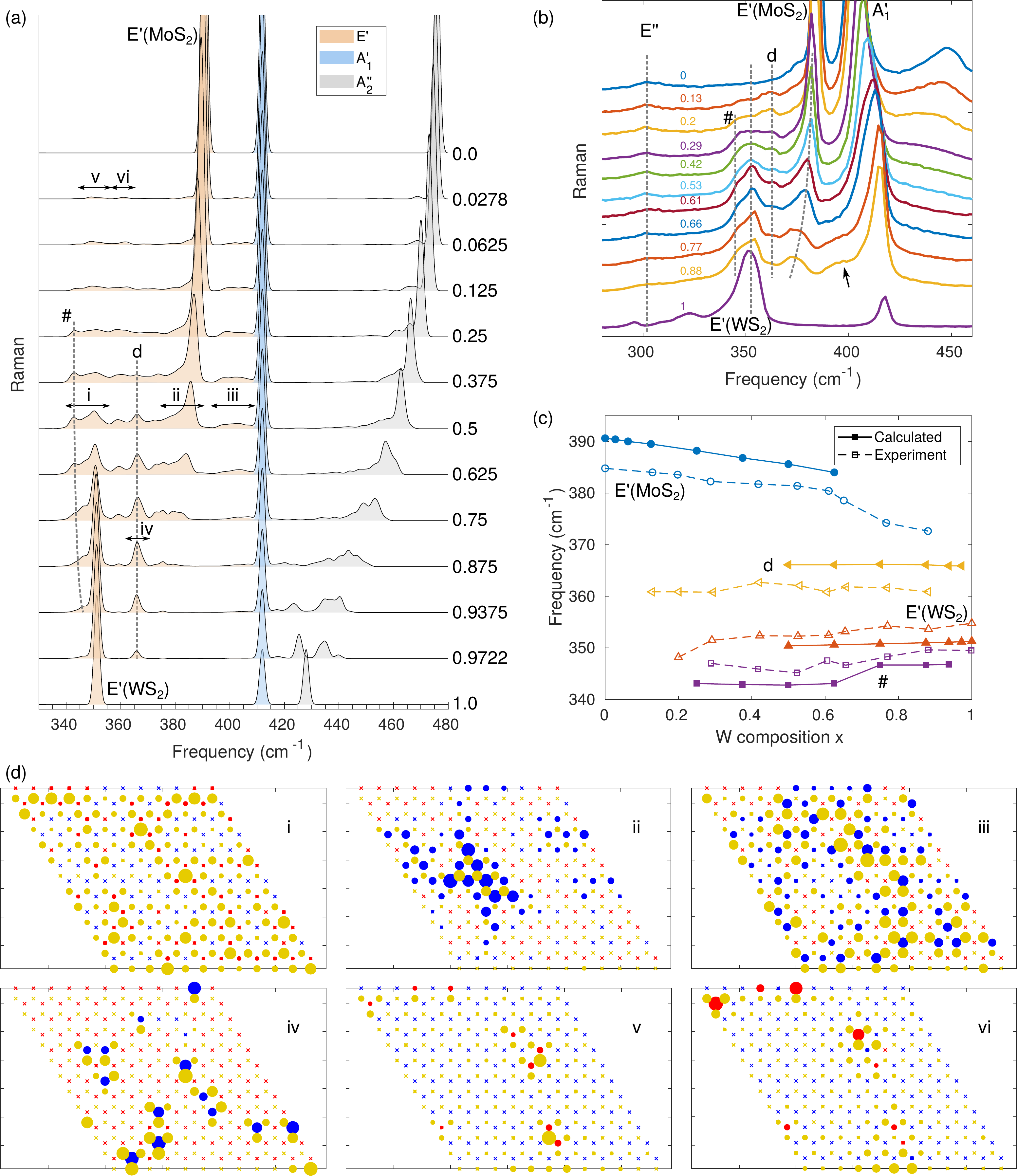}
\end{center}
\caption{\label{fig:MoWH}
(a) RGDOS for Mo$_{1-x}$W$_{x}$S$_2$ alloy, $x$ ranges from 0 to 1.
The total RGDOS is shown with solid black line.
The contributions from E$'$, A$_1'$, and A$_2''$ modes are shown by
yellow, blue, and grey shaded areas, repectively.
(b) Experiments, adapted from Ref.\ \onlinecite{Chen14_Nanos}.
(c) The calculated compositional dependence of the Raman peaks frequencies vs. the experimental counterparts.
(d) Illustration of selected eigenmodes (i - iv) from (a).
The blue, red, and yellow symbols correspond to Mo, W, and S atoms, respectively. 
The atoms are positioned in the supercell and the magnitude of symbols is propertional
to the amplitude of vibrations.}
\end{figure*}

The simulated Raman spectra for H-(Mo,W)S$_2$ monolayer as a function of
the composition are shown in Fig.\ \ref{fig:MoWH}(a), and
which can be compared to experimental Raman spectra shown
in Fig.\ \ref{fig:MoWH}(b) (from Ref.\ \onlinecite{Chen14_Nanos}).
The calculated A$_2''$ mode, although not Raman-active, is also shown, 
since it is infrared active and shows large changes with the composition.
To make it visible in the simulated spectra we use the same Raman tensor as for  A$_1'$.
The experimental and calculated peak positions are collected in Fig.\ \ref{fig:MoWH}(c).
The A$_1'$ mode consists of only chalcogen movement and thus in our
mass approximation approach this mode remains strictly constant.
Also E$''$ is unaffected by the MA and thus not shown in the calculated 
spectra, although its activation due to disorder is visible in 
the experimental spectra.

Overall a good agreement with the experiment is observed for the number of peaks as well as their positions:
(i) For the E$'$ mode, we confirm pronounced two-mode behavior
with the separate MoS$_2$- and WS$_2$-derived peaks. 
(ii) There is a clear downshift of the E$'$(MoS$_2$) peak, 
whereas the E$'$(WS$_2$) peak remains nearly constant in energy.
In experiment, at large W concentration the MoS$_2$-derived peak broadens
and possibly mixes with the d feature (marked d, as it was denoted
``disorder-related mode'' in Ref.\ \onlinecite{Chen14_Nanos}).
(iii) There are two additional features around the WS$_2$ peak:
one at about 345 cm$^{-1}$ (marked \#) and
one at about 360 cm$^{-1}$ (unmarked) in calculations.
The latter is difficult to observe in Fig.\ \ref{fig:MoWH}(b), but 
evident in the line shape fits in Ref.\ \onlinecite{Chen14_Nanos}.
(iv) Both in experiment and theory, at small W concentrations,
the W-derived features form a broad plateau below the E$'$(MoS$_2$) peak
with no particularly distinct peaks.
(v) A small bump develops between the E$'$(MoS$_2$)and A$_1'$ peaks,
which originates fully from the E$'$-mode.
While in calculations it prevails at intermediate concentrations,
in experiments this is only clearly visible at the W-rich side, 
and thus it is not clear if their origin is the same.

In order to understand the atomic origin of these peaks, we illustrate the eigenvectors from selected cases in Fig.\ \ref{fig:MoWH}(d),
where the sizes of the circles at the position of atom $k$ correspond to
the eigenvector weighted by the $\Gamma$-point projection
$|\vb(k0)|^2 \cdot w^2$ summed over all modes within the selected range of frequencies marked in Fig.\ \ref{fig:MoWH}(a).
As expected, the modes corresponding to MoS$_2$- and WS$_2$-derived peaks are
localized around Mo and W atoms, respectively.
The broader feature between E$'$ and A$_1'$ appears to be localized at the edges
of the Mo-regions (panel iii).
The ``disorder-related mode'' is not very visible at $x=0.5$, but at
$x=0.875$ our analysis clearly shows that it is localized to isolated Mo atoms
(panel iv).

The smaller peaks around it, on the other hand, are localized to
Mo-clusters (not shown), whose density at W-rich samples is naturally small.
The peaks denoted by $\#$-modes appear visually very similar to the main WS$_2$-derived modes and thus we think that this
shoulder just originates from asymmetrical broadening of the WS$_2$-peak.
On the other hand, this mode was assigned to 2LA(M) in Ref.\ \onlinecite{Chen14_Nanos}).
Our calculated LA(M) frequency for WS$_2$ is 177 \icm{}, yielding 2LA(M) at 354 \icm{},
and thus lies slightly above the E$'$(WS$_2$)-peak in our calculations, but could also
be slightly below the E$'$(WS$_2$)-peak in experiments.
Since we here only simulate the first-order Raman scattering, we know that the shoulder
in calculations contains no 2LA(M) contribution, but naturally we cannot
exclude such additional contribution in the experimental spectra.

\subsection{T'-(Mo,W)Te$_2$}

We next study T'-(Mo,W)Te$_2$ alloy,
which is computationally a significantly more challenging case,
since (i) the unit cell is larger and has lower symmetry than the H-phase,
thus leading to larger number of displacements in pristine system,
(ii) it is (semi-)metallic, necessitating the use of large k-point meshes.
The latter also means that the Raman spectra will necessarily be
resonant,
but the evaluation of the Raman tensor from the change of macroscopic
dielectric constant assumes non-resonant conditions.
Resonant Raman tensors can be used just as well in
our approach for simulated Raman spectra (Eq.\ \ref{eq:RGDOS}),
but their evaluation from first principles is again step up in computational
complexity and moreover makes the tensors frequency-dependent.
To avoid these problems, we here use the
non-resonant Raman tensors, which are moreover normalized in order to
better highlight all the Raman-active features, although this means
that the relative intensities of the peaks are not correctly captured.
The classification of the $\Gamma$-point vibrations, $\Gamma_{C_{i}}$ = 9 A$_{g}$ + 9 A$_{u}$,
shows that half (A$_g$) of the modes are Raman-active.
These modes can be arranged in two groups:
modes vibrating along the direction of the zigzag Mo/W chain, denoted by A$_g^{z}$,
and modes vibrating perpendicular to the zigzag chain, denoted by A$_g^{a}$.

\begin{figure}[!ht]
\begin{center}
\includegraphics[width=8.5cm]{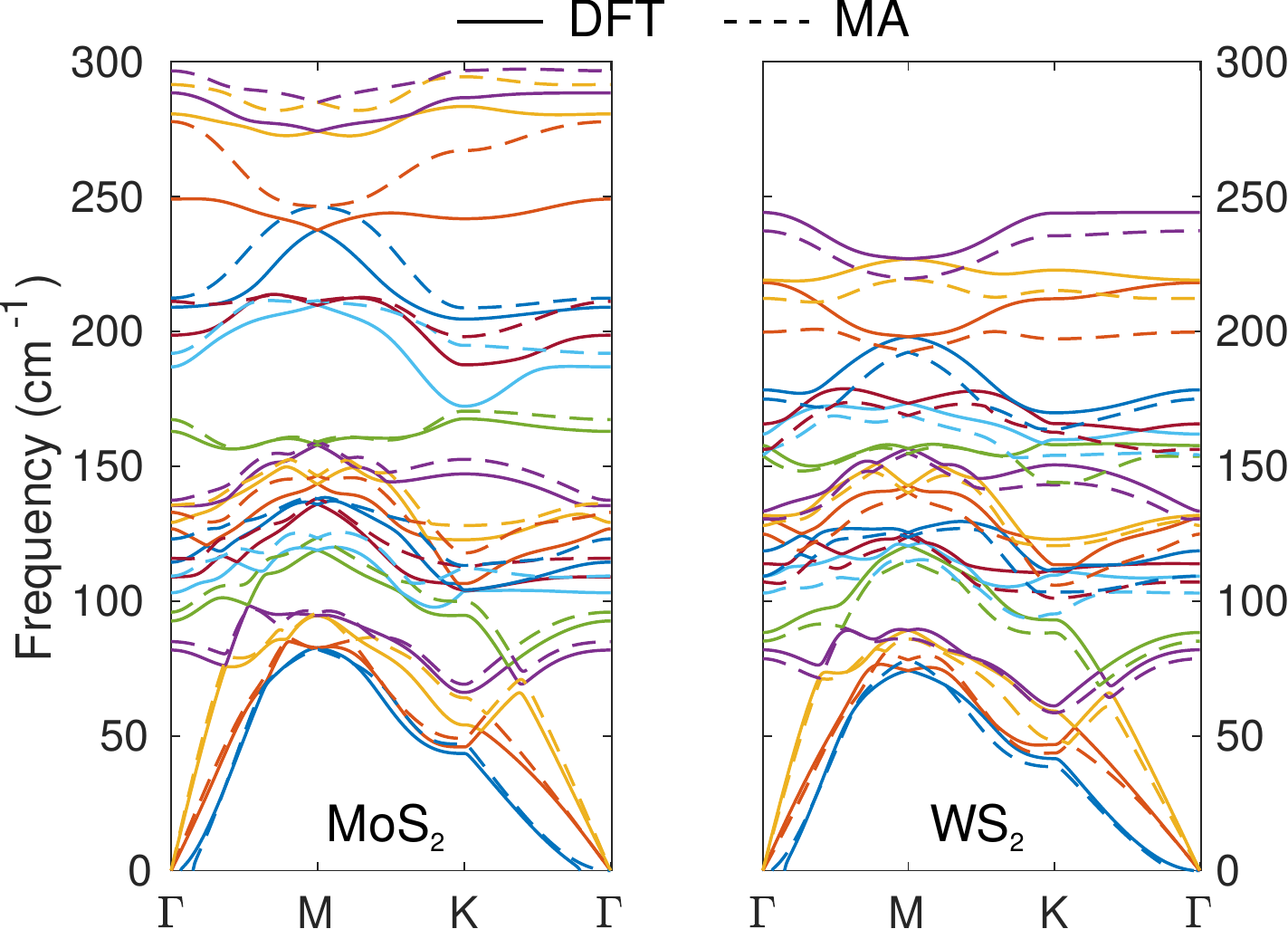}
\end{center}
\caption{\label{fig:massbandsTe}
Phonon dispersion curves of pristine MoTe$_2$ (left) and WTe$_2$ (right)
calculated either self-consistently using DFT (solid lines) or using
mass-approximation (dashed lines).
}
\end{figure}

Phonon dispersion curves calculated by DFT and by mass approximation are 
shown in Fig.\ \ref{fig:massbandsTe}. 
We again observe that frequencies from MA are shifted down by
about 10 cm$^{-1}$ in WTe$_2$, but the order and dispersion of the bands
is captured well.
The only clear deviation occurs for WTe$_2$ around 220 cm$^{-1}$
at the $\Gamma$-point,
where the quasi-degenerate Raman active modes from the DFT calculation
breaks into two modes at 200 cm$^{-1}$ and 212 cm$^{-1}$ from the MA
calculation,
echoing the splitting observed in MoTe$_2$ at 250 cm$^{-1}$ and 280 cm$^{-1}$.
This feature is observed in experiment for bulk (Mo,W)Te$_2$ \cite{Joshi2016}. 
It is worth noting that the lattice constants of MoTe$_2$ (3.37 \AA, 7.15 \AA)
and WTe$_2$ (3.42 \AA, 7.12\AA) are not quite as close as
those of the parent compounds in H-(Mo,W)S$_2$.

\begin{figure*}[ht!]
\begin{center}
\includegraphics[width=17cm]{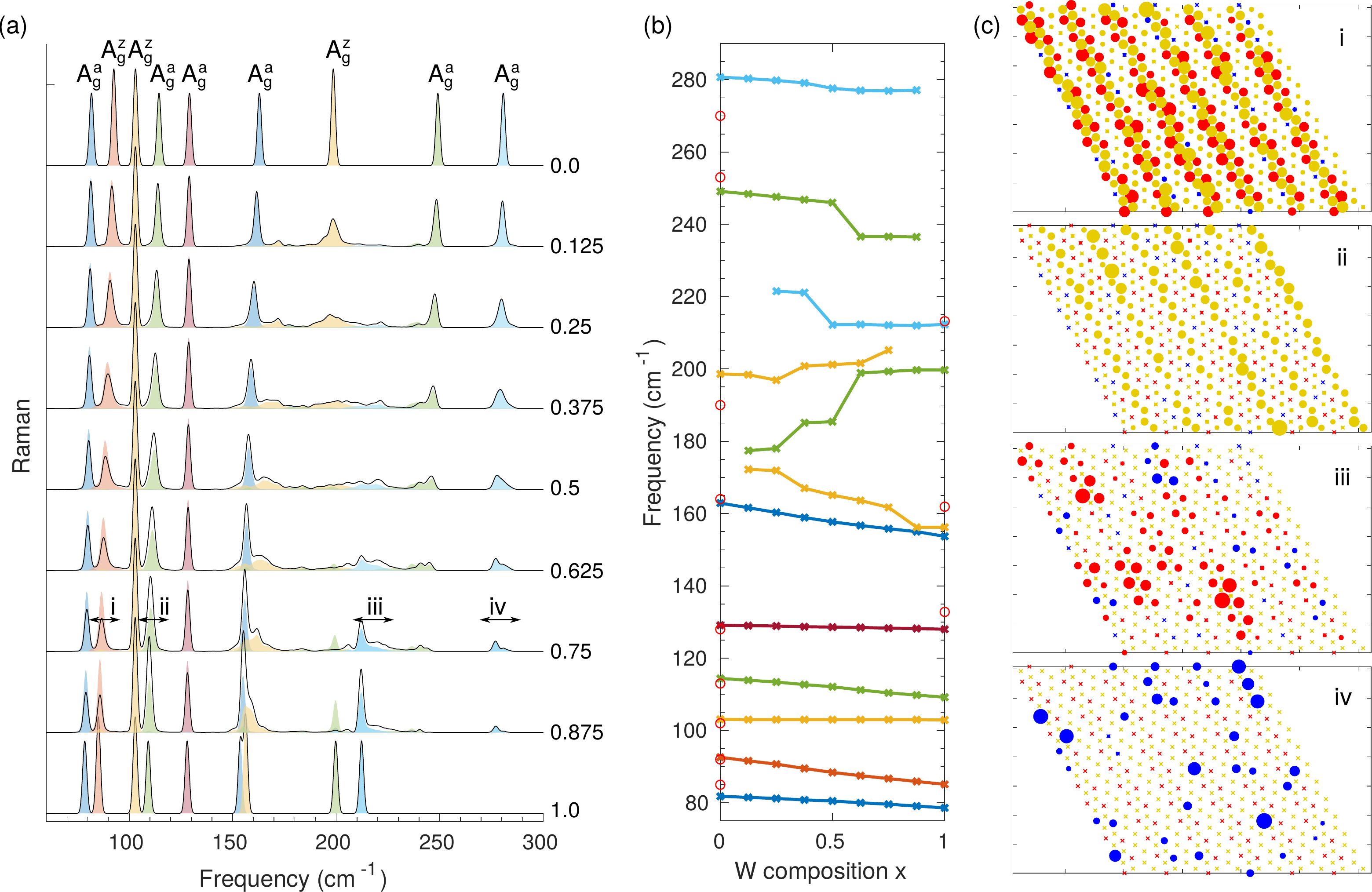}
\end{center}
\caption{\label{fig:MoWTe}
(a) RGDOS for T'-(Mo,W)Te$_2$.
The total RGDOS is shown with solid black line.
The shaded areas show contributions from the projection to
eigenmodes of the pristine T$'$-MoTe$_2$.
The modes are colored sequentially (and loops once).
(b) Evolution of the peak maxima positions for Mo$_{1-x}$W$_x$Te$_2$ alloys.
Experimental data (red open circles) are taken from Refs.\ \cite{Jiang16_SRep,Chen17_ACSNano}.
(c) Illustration of selected SC eigenmodes from the $x=0.75$ case, as indicated in (a).
The SC eigenmodes within the frequency range are weighted by the
projection to the dominant PC eigenmode.
The blue, red, and yellow symbols correspond to Mo, W, and Te atoms, respectively.
}
\end{figure*}
The calculated RGDOS for the monolayer T'-(Mo,W)Te$_2$ alloy 
as a function of composition are shown in Fig.\ \ref{fig:MoWTe}(a)
and the peak positions are collected in Fig.\ \ref{fig:MoWTe}(b).
We remind, that while in H-(Mo,W)S$_2$ the alloy modes could be easily assigned 
to the pristine modes from which they originated
thanks to the large separation in frequency, 
here due to the large number of modes, the mixing is more complicated 
and thus the eigenmode-projection is necessary to distinguish between 
the Raman-active and -inactive features.
The projection scheme allows us to distinguish the origins of each
peak in terms of the primitive cell eigenmodes, revealing
that the ordering of the modes is retained in the same order 
throughout the alloys.
The eigenvectors of these modes in the parent phases
have been illustrated in several previous works
\cite{Jiang16_SRep,Kim16_Nanos,Beams16_ACSNano,Chen16_NL,Grzeszczyk16_2DM,Zhang16_NComm,Wang17_AFM,Chen17_ACSNano}, and are not repeated here.
Nevertheless, they show that the six lowest frequency modes are mostly
localized to Te atoms, and the three high frequency modes to Mo/W atoms. 
Consequently, the six lowest frequency modes exhibit
single-mode behavior and the three high frequency modes
two-mode behavior, reflecting the fact that alloying is carried
out in the metal sublattice.
Among the six lowest frequency modes that exhibit the single mode behavior,
the third one is silent in the metal sublattice and the fifth one nearly silent \cite{Chen17_ACSNano},
and thus they show very little changes upon alloying.
There are also clear differences in the degree of the alloying-induced broadening of the other four peaks,
with the first one showing least broadening,
the second one the strongest broadening,
and the fourth and sixth modes falling in between.
Fig.\ \ref{fig:MoWTe}(c) illustrates the second and fourth modes of the 
x=0.75 alloy. The fourth mode (panel ii) is localized very clearly only on the 
Te atoms and mostly on the rows with long metal-metal distance,
whereas the second mode has also weight on the metal atoms and is mostly
localized on the rows with short metal-metal distance.

The last three modes in Fig.\ \ref{fig:MoWTe}(a) show a very clear 
two-mode behavior with splitting into MoTe$_2$ and WTe$_2$-like modes at intermediate alloy concentrations.
The eigenvectors in Fig.\ \ref{fig:MoWTe}(c) show that these modes are localized
almost completely on the metal atoms and the two-mode behavior reflects the localization around Mo and W atoms.
The eigenmode projections illustrated in Fig.\ \ref{fig:MoWTe}(c) are found to provide additional insight into the peak origins.
For instance, 
there is a mode at 200 \icm{} in both the MoTe$_2$ and WTe$_2$ phases,
but the projections reveal that they correspond to different modes.
Somewhat similarly, the 160 \icm{} peak in WTe$_2$ is seen to contain two modes,
which in the MoTe$_2$ region are located at 160 \icm{} and 200 \icm{}.

Comparison to experimental results is hindered by the fact, that
to the best of our knowledge, all the experimental
T'-Mo$_{(1-x)}$W$_x$Te$_2$ alloy results are from bulk samples
\cite{Revolinsky64_JAP,Oliver17_2DM,Lv17_SRep,Rhodes17_NL}.
Monolayer data is only available for pure MoTe$_2$ and WTe$_2$
\cite{Chen17_ACSNano,Jiang16_SRep,Kim16_Nanos}.
Naturally, there exists also a large body of data for pure bulk or few-layer phases
\cite{Joshi16_APL,Beams16_ACSNano,Ma16_PRB,Chen16_NL,Grzeszczyk16_2DM,Wang17_AFM,Zhang16_NComm}.
Although the bulk and monolayer frequencies are generally fairly close,
to facilitate a proper comparison, in Fig.\ \ref{fig:MoWTe}(b) we only show the available 
monolayer results for MoTe$_2$ and WTe$_2$.
For the low-frequency modes in MoTe$_2$ and WTe$_2$, calculated and experimental 
frequencies agree very well.
The agreement deteriorates for high-frequency modes, 
but the experimental and calculated peaks can still be mapped.
Also the ordering of the 
A$_g^{a}$ and A$_g^{z}$ modes is correctly reproduced.
When comparing to the bulk alloy results, 
our calculations indicate that the reported disorder-activated modes around 180 cm$^{-1}$
and 202 cm$^{-1}$ \cite{Oliver17_2DM},
can be a mix of the last three high-frequency modes and can be tuned by varying composition.
Our calculations produce a large number of small peaks at these frequencies, with contributions
from all the three high-frequency modes, but we do not obtain one or two prominent
peaks. This might be caused by normalization of Raman tensors
in our simulated spectra.
The peak at 130 \icm{} in MoTe$_2$ was found to split into two peaks separated by about 3 \icm{}
upon increasing the W concentration \cite{Oliver17_2DM,Lv17_SRep},
and was assigned to mixing in Ref.\ \onlinecite{Oliver17_2DM} and 
to a phase change from monoclinic to orthorhombic lattice in Refs.\ \cite{Lv17_SRep,Chen16_NL}.
Since this peak is silent in the metal sublattice, it shows no alloying-induced splitting
nor even any broadening in our calculations, and thus our calculations do not
support the assignment to mixing.
For the highest frequency mode, our calculations correctly capture the broadening toward
higher frequencies on both the MoTe$_2$ and WTe$_2$ regions \cite{Oliver17_2DM}.

\subsection{Impurities in H-MoS$_2$}

The Raman signatures can be used to identify impurities
at small concentrations (small with respect to alloying, i.e.,
within few percent).
In some instances, as seen also in the previous sections,
impurities can produce very distinct new peaks,
broaden existing peaks, or result in very broad features.
In this section, 
we insert a small number of impurity atoms into the lattice
and examine the trends in the changes of the Raman spectra.
The mass approximation limits our study
to cases where chemical bonding upon substitution is expected to remain
fairly similar.
To this end, 
we either replace the Mo atom by other transition metal element
or the S atom by an atom from the nitrogen, oxygen, or fluorine groups.
Clearly, this is expected to work best for the elements in the same column
in the periodic table and worsen the further away from it.
The small impurity concentration helps to avoid problems
with the large strain.
For the calculations, 
we here adopt a slightly simplified procedure, 
where we simply take the 5$\times$5 supercell with a single impurity.
This is sufficiently large to describe the localized modes,
and while the peak broadenings would not be correctly described, there are
very little changes in the position and broadening of the main
peaks in these dilute cases.

\begin{figure*}[!ht]
\begin{center}
\includegraphics[width=\textwidth]{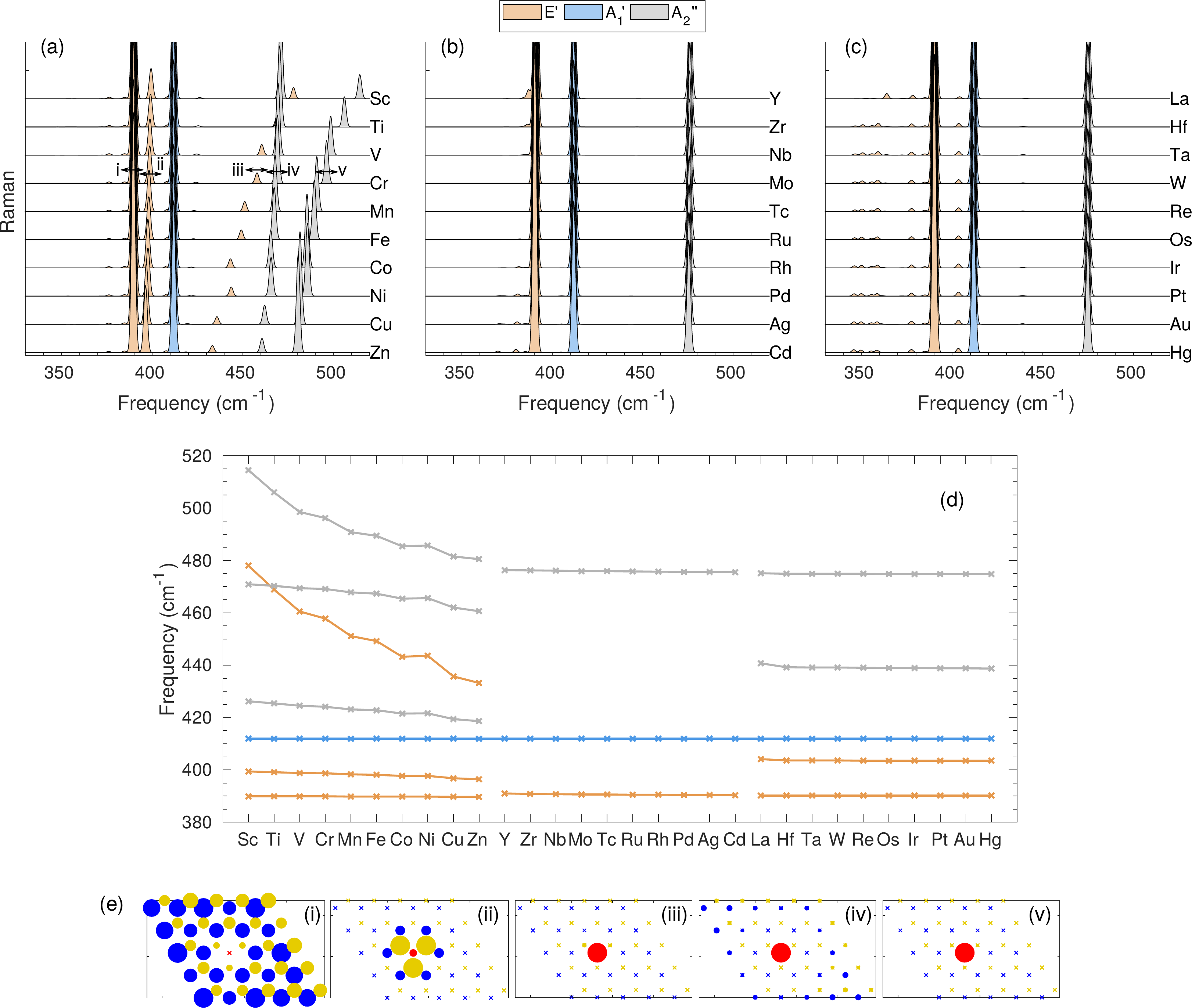}
\end{center}
\caption{\label{fig:impuM}
(a-c) RGDOS for impurities in the Mo site in MoS$_2$, grouped by the rows in
the periodic table.
(d) Positions of the peak maxima extracted from panels (a-c).
(e) Selected eigenmodes from the Cr case.
The blue, red, and yellow symbols correspond to Mo, Cr, and S atoms, respectively.
}
\end{figure*}

The RGDOS for the Mo-site impurities are shown in Fig.\ \ref{fig:impuM}(a-c).
One impurity in 25 lattice sites corresponds to the $4 \%$ impurity concentration.
The behavior is clearly different for 3d, 4d, and 5d transitional metal impurities. Following the impurity masses, the additional
impurity induced peaks are at highest frequencies for the 3d elements
and at lowest frequencies for 5d elements, whereas the 4d impurities
show very little new features.
In case of the 3d elements, there is a pronounced splitting between
the E$'$ and A$_2''$ modes
and an additional, mostly E$'$-derived, mode between the two.
We note again that A$_2''$ mode is not Raman-active, and only shown here for reference.
The eigenmodes are shown in Fig.\ \ref{fig:impuM}(e).
Not surprisingly, the main peak is localized in the MoS$_2$ regions
(panel i)
The second E$'$ feature is localized around the impurity (panel ii)
and the last one strictly at the impurity (panel iii).
This last E$'$ peak should have appreciable Raman intensity and 
frequency that sensitively depends on the transition metal impurity
and thus seems to provide the most effective impurity signature.
For the two A$_2''$-derived peaks, the lower frequency mode is localized in the MoS$_2$ regions (panel iv, the Cr atom shows intense due to its small mass,
but all Mo atoms are also active) and the higher frequency one around 
the impurity (panel v).

Very little happens with the 4d impurities, only a small shift of the main
E$'$ mode together with slight broadening,
stemming from the small (relative) change of the mass. 
All the 5d impurities show features similar to the (Mo,W)S$_2$ alloy considered previously: a broad set of weak features at 350--400 \icm{}
and one peak between E$'$ and A$_1'$ peaks.
For the two eigenmodes shown in Fig.\ \ref{fig:MoWH}(d) (panels v,vi),
despite having clearly different frequencies, they have fairly similar eigenvectors.
Since the MoS$_2$ E$''$, A$_1$, and A$_2''$ modes at the K and M points
largely fall at frequencies between 350 and 400 cm$^{-1}$,
we think these impurity modes have large contributions from the 
off-$\Gamma$ k-points and only a small $\Gamma$-point,
Raman-active contribution.
In essence, these impurities lead to mixing of the vibrational modes
at different q-points of the primitive cell BZ.
No pronounced features are observed at low frequencies,
and there are no gap states.

Overall, it appears that it should be possible to resolve the presence of even
fairly dilute concentration of 3d transition metal impurities in MoS$_2$ 
from the splitting of the E$'$ peak, possibly even with the elemental precision,
although the absolute values given here may suffer from the limitations 
of the mass approximation.
Dilute concentration 4d impurities are expected to be largely invisible in Raman,
whereas 5d impurities might show up in Raman but their identification can be
difficult.

\begin{figure}[!ht]
\begin{center}
\includegraphics[width=8.5cm]{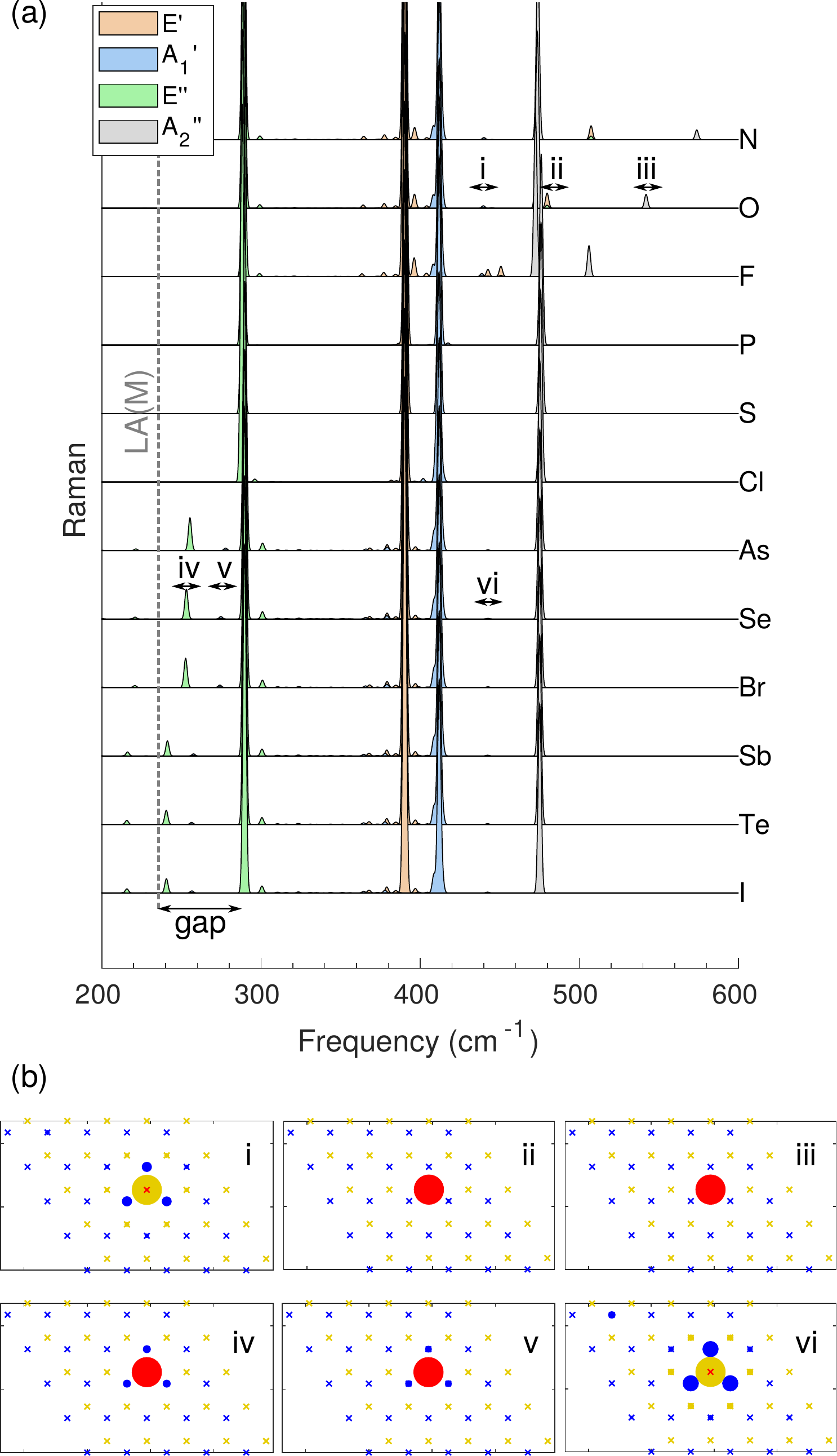}
\end{center}
\caption{\label{fig:impu}
(a) RGDOS for impurities in the S site in MoS$_2$.
The LA(M) frequency from pristine MoS$_2$ and the gap in the phonon structure
are also indicated.
(b) Selected eigenmodes for the O and Se impurity systems.
The blue, red, and yellow symbols correspond to Mo, O/Se, and S atoms, respectively.
}
\end{figure}

The RGDOS for the S-site impurities in MoS$_2$ are shown in Fig.\ \ref{fig:impu}(a).
One impurity in 50 lattice sites corresponds to $2 \%$ impurity concentration.
Again, lighter impurities lead to additional peaks at higher frequencies
and heavier impurities at lower frequencies, but the features that are
most likely to be observed in experiments are those falling 
above the A$_2''$ mode or inside the gap between E$''$ mode and the LA(M) edge.
In fact, such features have been reported in the literature for MoS$_2$ with
light Se alloying at about 270 cm$^{-1}$ \cite{Li14_JACS,Su14_FER,Mann14_AM,Feng15_ACSNano}
and with light Te alloying at about 243 cm$^{-1}$ \cite{Yin18_Nanot}, 
agreeing well with our calculations.

O and Se impurities in MoS$_2$ are chosen as representative examples to be discussed in more detail.
Selected eigenvectors of these impurity systems are presented in Fig.\ \ref{fig:impu}(b).
In the case of the O impurity, the feature (ii) just above A$_2''$ is mainly derived from E$'$ with a small E$''$ contribution and
it should thus be visible in Raman measurements.
The high frequency feature (iii) is mostly of A$_2''$ type, but it contains also
an appreciable A$_1'$ contribution and thus could also be visible.
In the case of the Se impurity, there are two features in the gap,
with the lower one (iv) derived mostly out of E$''$ with some E$'$,
and the higher one (v) mostly from the pristine A$_1'$ mode with some A$_2''$ character.
Finally, we mention the features (i) and (vi), which
are localized mostly at the S atom on the opposite side of the layer from 
the impurity atom and thus they also have the same frequency, independent 
of the impurity element.
While this feature is barely visible in the simulated spectrum, it is derived mostly 
from the pristine A$_1'$ mode and thus could be observable.

\section{Conclusions}

We have devised an efficient computational method to simulate Raman spectra 
of large systems, being
especially applicable to alloys and also systems with small number of defects.
The method is based on the projection of vibrational eigenvectors of
the supercell to the eigenvectors from the primitive cell and using
them as weights in summing over the Raman tensors calculated at the primitive cell.
We moreover used mass approximation to rapidly evaluate the vibrational
modes in the supercell.
We applied the method to two different transition metal dichalcogenide
monolayer alloys, H-(Mo,W)S$_2$ and T'-(Mo,W)Te$_2$, 
and to impurities in H-MoS$_2$.
The accuracy of the method was validated in the case of H-(Mo,W)S$_2$ 
alloy through comparison to the available experimental reports.
T'-(Mo,W)Te$_2$ and impurity cases are used to 
(i) demonstrate the wider applicability of the method and
(ii) provide predictions in few technologically relevant systems.
We note that in addition to yielding the simulated Raman spectra,
the projection scheme also provides a powerful tool for analyzing
the origin of the Raman-active features.
The method presented here is not limited to 2D materials, and is 
applicable to various other bulk and low-dimensional systems.

\section*{Acknowledgments}
We thank Prof. Liming Xie for providing us the experimental data.
We are grateful to the Academy of Finland for the support under Projects No.~286279 and 311058.
We also thank CSC--IT Center for Science Ltd. and
Aalto Science-IT project for generous grants of computer time.


%

\end{document}